\documentclass[runningheads]{llncs}
\usepackage{graphicx}
\usepackage{times}
\usepackage{grffile}
\usepackage{amsmath}
\usepackage{amssymb}
\usepackage[breaklinks=true,bookmarks=false]{hyperref}
\graphicspath{{pics/}}
\begin{document}
\title{Photometric light curves classification with machine learning}
%
%
\author{Tatiana Gabruseva\inst{1} \and
Sergey Zlobin\inst{2} \and
Peter Wang\inst{3}
}
\authorrunning{T. Gabruseva, S. Zlobin and P. Wang}
%
\institute{Independent researcher, Ireland - Croatia - Montenegro \and
ABBYY Production, Moscow, Russia \and
Monash University, Clayton, Victoria, Australia\\
\email{tatigabru@gmail.com}\\
\email{}}
\maketitle              
\begin{abstract}
The Large Synoptic Survey Telescope will complete its survey in $2022$ and produce terabytes of imaging data each night. To work with this massive onset of data, automated algorithms to classify astronomical light curves are crucial.

Here, we present a method for automated classification of photometric light curves for a range of astronomical objects. Our approach is based on the gradient boosting of decision trees, feature extraction and selection, and augmentation. The solution was developed in the context of The Photometric LSST Astronomical Time Series Classification Challenge (PLAsTiCC) and achieved one of the top results in the challenge.
\end{abstract}
%
%
\section{Introduction}
\noindent The Large Synoptic Survey Telescope (LSST) will achieve first light in $2019$ and commence its 10-year survey in $2022$~\cite{lsst}. LSST data are beneficial for a range of scientific studies, such as studying stars in our Galaxy, understanding how solar systems formed, probing nuclear physics with brief burps before stars explode, understanding the role massive stars play in shaping their chemistry, measuring how much matter there is in our Universe, and others.

Over ten years, LSST will make a slow-motion movie of half the sky, visiting each location roughly twice per week. LSST will produce $15$ Terabytes of imaging data, with up to $\approx10^7$ transient detections each night~\cite{Kessler2019}. Every week LSST will find more sources that vary with time than Hubble has ever seen in its entire 28+ year life~\cite{starter}. To deal with this massive onset of data, automated algorithms to classify astronomical light curves are crucial~\cite{dataset}.

Recently, the machine learning approaches have demonstrated enormous potential for classification of variable stars~\cite{Richards2011} and transients \cite{Lochner2016,Karpenka2012}. The Photometric LSST Astronomical Time-Series Classification Challenge (PLAsTiCC) is an open challenge hosted on kaggle platform with the task to classify simulated photometric light curves for the range of astronomical objects in preparation for observations from the LSST surveys~\cite{kaggle,plasticc_arxiv}.

Here we present the solution of the PLAsTiCC Classification Challenge. Our approach is based on data augmentation, feature extraction and selection, and classification with the gradient boosting of decision trees. The method achieved one of the best results for a single model performance in the challenge.

\section{Dataset}
The training dataset consisted of simulated astronomical light curves modeled for a range of transients and periodic objects~\cite{dataset}, see~\cite{Kessler2019} for the details on the dataset simulation. The description of the classes present in the training set and their distribution is in Table 1. The training dataset had $7848$ light curves and was highly unbalanced.
\begin{table}[htb]
\begin{center}
\begin{tabular}{|l|c|c|l|}
\hline
Class Id & DDF samples & WFD samples & Description  \\
\hline
90& 837& 1476&SNIa WD detonation, Type Ia \\
67& 66& 142&SNIa-91bg Peculiar Type Ia\\
52& 70& 113&Peculiar SNIax \\
42& 304& 889&Core collapse, Type II SN \\
62& 129& 355&Core collapse, Type Ibc SN \\
95& 41& 134&Super-lum. SN (magnetar) \\
15& 8& 487&Tidal disruption event \\
64& 2& 100&Kilonova \\
88& 119& 251&Active galactic nuclei \\
92& 58& 181&RR Lyrae \\
65& 309& 672&M-dwarf stellar flare \\
16& 162& 762&Eclipsing binary stars \\
53& 4& 26&Mira, Pulsating variable stars \\
6 & 7& 144&Single $\mu$-lens from a single lens \\
\hline
\end{tabular}
\end{center}
\label{table:classes}
\caption{The description of the classes present in the training dataset and their distribution}
\end{table}
The PLAsTiCC challenge included two of the LSST observing modes, namely: wide-fast-deep (WFD) and deep-drilling-fields (DDF). The WDF is the primary component of LSST observations strategy; it covers almost half the sky with sub-optimal temporal resolution. The DDF is a specialized mini-survey with a set of $5$ telescopes covering almost 50 deg of the sky. The DDF observations are $20$ times more frequent compared to WFD with the same exposure time, they provide much more defined light curves; however, at the cost of the sky observation region. Both observation methods were added within the same night in the challenge, with DDF nightly visits being $\approx2.5$ more frequent and $\approx1.5$ mag deeper compared to WFD.

\begin{figure}
\includegraphics[width=1\textwidth]{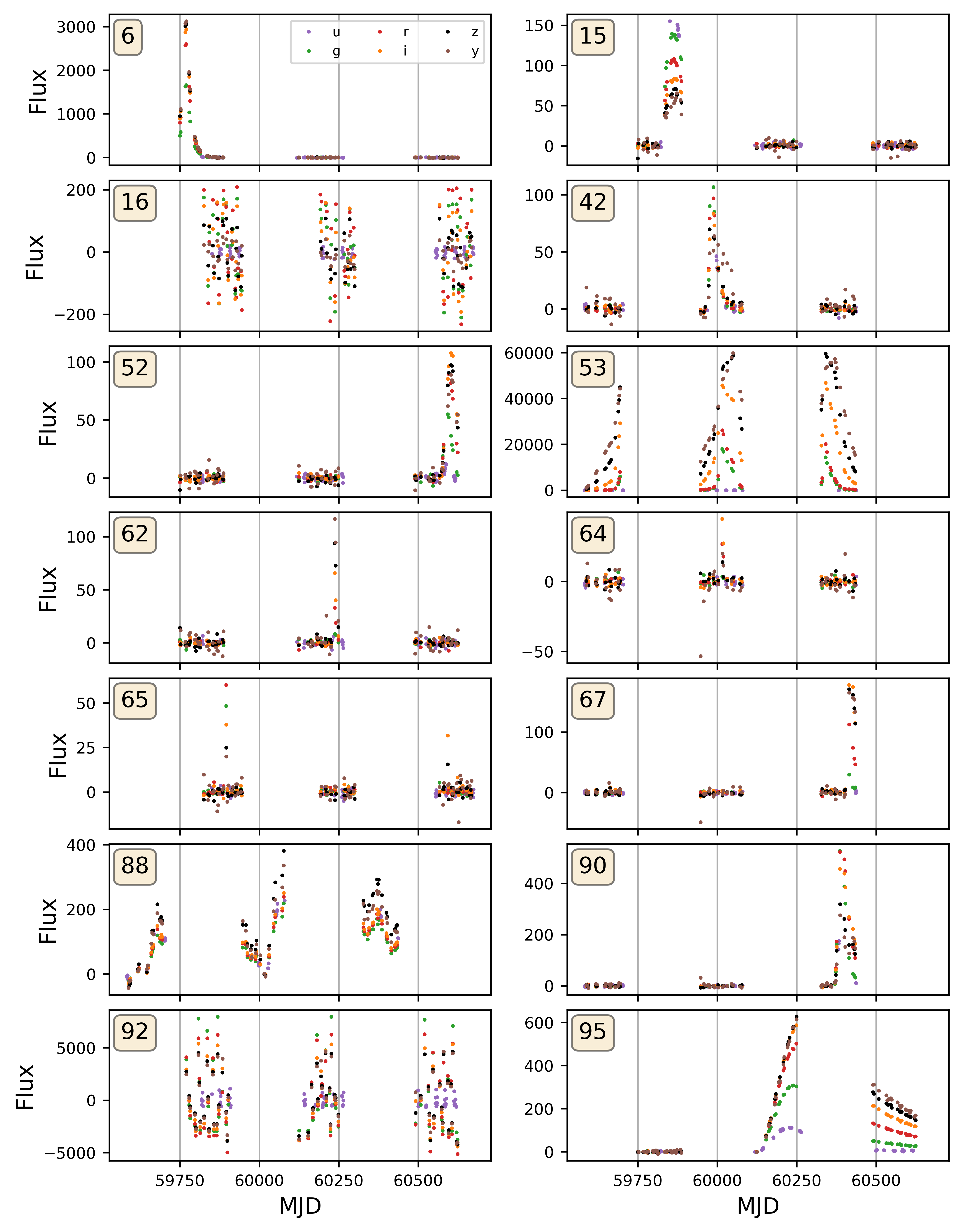}
\caption{Examples of simulated light curves for each class for DDF mode in the passbands $ugrizy$. MDJ -- Modified Julian Date in days} \label{fig:ddf}
\end{figure}

\begin{figure}
\includegraphics[width=1\textwidth]{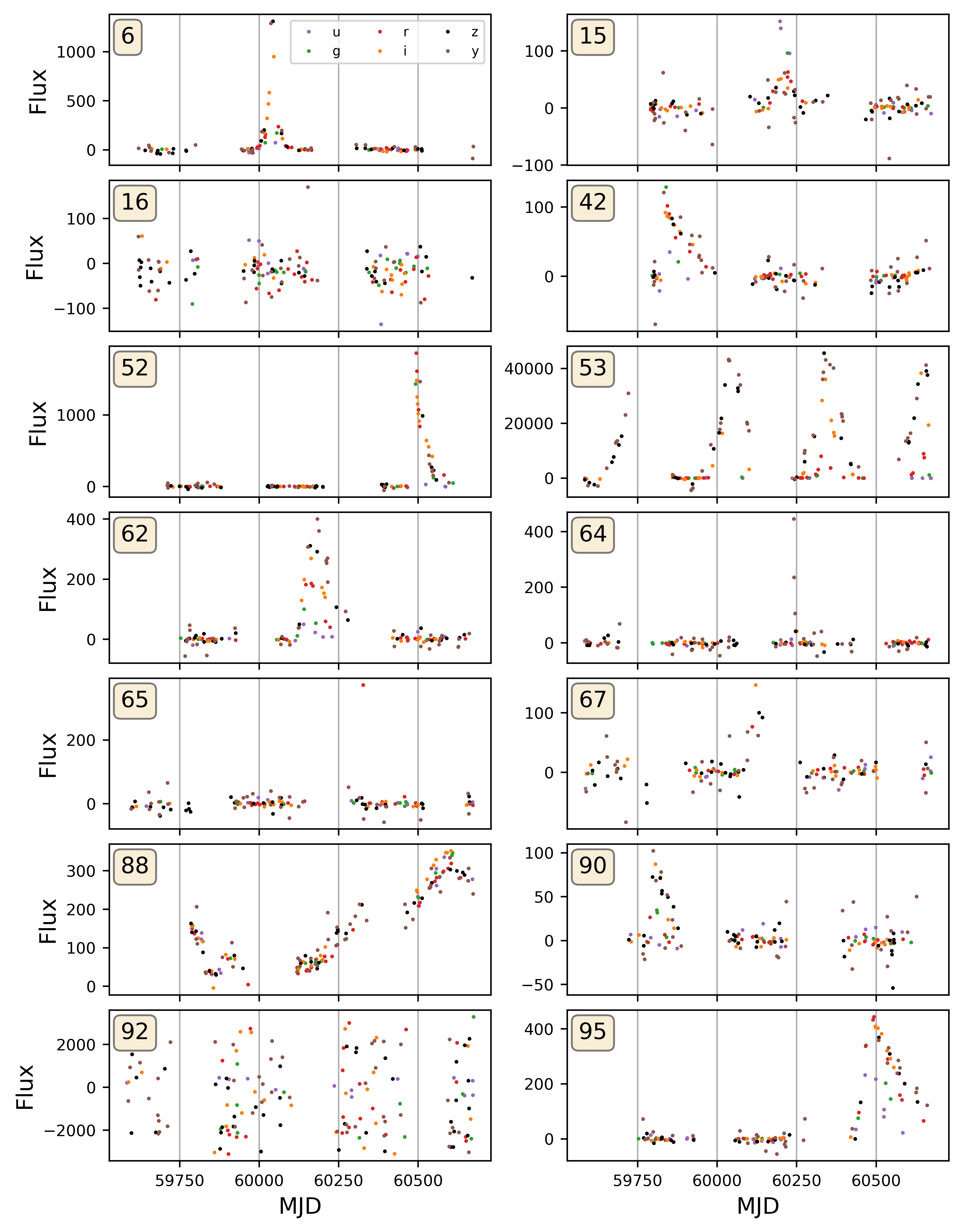}
\caption{Examples of simulated light curves for each class for WFD mode in the passbands $ugrizy$.} \label{fig:wfd}
\end{figure}

The light curves were measured in $6$ spectral passbands $ugrizy$~\cite{starter}. Simulated light curves for DDF and WFD observation modes are shown in Fig.~\ref{fig:ddf} and Fig.~\ref{fig:wfd}, respectively, for all classes. More examples can be found in~\cite{eda_kernel}.

Besides \textit{flux} values for all 6 passbands, the dataset included \textit{flux\_err} values that simulate measurement errors of the flux (at the $68\%$ confidence interval of the measurement). For extragalactic classes data included the photometric redshifts of the host-galaxies, \textit{hostgal-photoz}, with their uncertainty, \textit{hostgal-photoz-error}, the  distance modulus, \textit{distmod}, and sometimes spectroscopic redshifts, \textit{hostgal-specz}, if available. Each object was provided with its galactic coordinates, right ascension and the declination, but they did not contribute to the model.

Also, each point had a boolean \textit{detected} flag. LSST uses a reference image for each position on the sky, a template, and measures the astrophysical sources against the template. Objects for which the total flux from source and background is higher than that of the template have positive flux, and when the total flux is lower than the template flux, they have a negative flux. Due to this difference imaging techniques, objects have to have changed significantly with respect to the template to be detected by LSST. Whether or not they were detected with respect to the template is reported as \textit{detected=1}. For more details on the metadata provided, see~\cite{starter}.

\section{Evaluation}
The models were evaluated with weighted multi-class logarithmic loss. Such an arrangement allowed for each class to have roughly equal contribution to the evaluation score.
Each object has been labeled with one type. For each object one get a set of predicted probabilities for it to belong to a particular class. The formula used for the evaluation is then defined as follows:
\begin{equation}
\text{Log Loss} = - \left( \frac{\sum^{M}_{i=1} w_{i} \cdot \sum_{j=1}^{N_{i}} \frac{y_{ij}}{N_{i}} \cdot \ln  p_{ij} }{\sum^{M}_{i=1} w_{i}} \right),
\end{equation}
where $N$ is the number of objects in the class set, $M$ is the number of classes,  $y_{ij}$ is $1$ if observation $i$ belongs to class $j$ and $0$ otherwise, $p_{ij}$ is the predicted probability that observation $i$ belongs to class $j$.

In order to avoid the extremes of the log function, predicted probabilities values are clipped to:
\begin{equation}
\max(\min(p,1-10^{-15}),10^{-15}).
\end{equation}

\section{Features extraction from photometric light curves}
This section details the features extracted to represent light curves. As telescope observations are not always taken at uniformly spaced intervals, the light curve features should be invariant to this non-uniformity.

\subsubsection{Statistics}
We calculated basic statistics from the light curves and tested all features from the cesium package \cite{cesium}. The features based on ratios, $median\_absolute\_deviation$, standard deviation ($std$), and skewness ($skew$) were the most useful for classification. The time between the first and the last detected observation ($detected\_time\_diff$), suggested by Dr Grzegorz Sionkowski, is a handy feature helping to separate transient supernova events.
The normalized maximum flux values, $norm\_flux\_max\_i$, defined as $max(flux\_i)/max(flux)$, where $i$ denotes the passaband number, also shown high importance for the classifier.

\subsubsection{Light curves features from the FEETS package}
There is a number of packages for feature extraction from time-series. We tested features from the FEETS package, specially designed for the photometric light curves~\cite{feets}. $CAR\_sigma$, $CAR\_tau$, and $CAR\_mean$~\cite{Pichara2012} were particularly helpful, however, required long calculation time. We used the numba package \cite{numba} available for python along with minor vectorisation to speed up their calculation $10$x times faster than the original package.

\subsubsection{Peaks analysis}
For the most well-defined passbands 2, 3, 4, and 5 we look for a single peak if it exists. A peak is searched by a threshold relative to the flux maximum of a passband. We consider the certain peak intervals (with the known flux values) and estimated peak intervals (by linear approximation of flux values). Then we get the aggregated “certain” peak interval by the union of passband “certain” intervals, and wider aggregated “estimated” interval by the intersection of passband “estimated” intervals. The lengths of these two aggregated intervals, $peaks\_certain\_width$ and $peaks\_estimated\_width$, are used as features.

Another common way for peaks analysis is an autocorrelation and cross-correlation. They require equal temporal sampling; therefore, data for the light curves were first re-sampled using linear interpolation, and then the parameters of the autocorrelations were extracted. We used autocorrelation lag $0$ ($autocorrelation$) and autocorrelation peak widths. We tried to extract statistics from the interpolated and re-sampled light curves as well, but it did not bring the model improvement.

\subsubsection{Magnitudes}
The magnitudes, $M$, were calculated from the measured flux ($F$) and distance modulus, $\mu$, as follows:
\begin{equation}
m = -2.5*log_{10}(\frac{F}{F_{ref}}) - \mu,
\end{equation}
where $F_{ref}$ is a reference flux. Brighter objects have smaller magnitudes. For the purpose of the challenge, the reference flux was set to $F_{ref} = 1$ for all the objects as the reference was not provided.

We tested two sets of magnitude features: calculated for all data points and data points with $detected = 1$ only. The latter option gave better results, possibly due to better handling of outliers in the observations. We calculated statistics of magnitudes per passband ($magnitude\_min\_i$, $i=0..5$) and aggregated for all passbands together. The magnitude minimum, $magnitude\_min$, mean, $magnitude\_mean$, standard deviation, $magnitude\_std$, and also integration over the peak, $magnitude\_sum$ shown the highest importance for the classifier (see Table 3). The statistics calculated for individual passbands had a smaller contribution to the classification model accuracy than the aggregated values.
The magnitudes were also corrected for the redshift using the following k-correction calculator~\cite{kcorrection}. It is only valid for relatively small redshifts~\cite{Beare2014}, but the majority of data lie in this region. The k-correction of magnitudes per band did not improve the classification results.

\subsubsection{Parametric curve fitting}
The main challenge was to separate supernova of different types, classes: 42, 52, 62, 67, and 90. The other classes were well distinguished with the basic features listed above, therefore we focused on the special features to differentiate between supernova types.
The parametric curve fitting is a commonly used technique for supernova classification~\cite{Bazin2009,Newling2011,Karpenka2012,Gonzalez-Gaitan2015}.

We tried models from the papers of Bazin et.al.~\cite{Bazin2009}, Newling et.al~\cite{Newling2011}, and Karpenka et.al.~\cite{Karpenka2012} for features extraction. Besides that, we also calculated the Gaussian fits to define peak-width, and linear decay slope of supernova light curves in the log scale for different passbands, $supernova\_decay\_i$. The root-mean-square loss was one of the most important features in all the curves fittings.

The parametric curve fitting was made in two stages. First, we fit a Gaussian model to a light curve to define an approximate position of the light curve peak:
\begin{equation}
f^{(i)}(t)=f_m^{(i)} \cdot e^{-\dfrac{(t-t_0)^2}{2s^2}} + f_0,
\end{equation}
where $i = 0..5$ is the passband number.
The Gaussian width $gauss\_s\_i$ and a root-mean-square fit error, $gauss\_error\_i$, were used as the classifier features.

We defined a penalty function to regularise $t_0$, so that band peaks do not wander too far off each other for different passbands:
\begin{equation}
p(\mathbf{t_0}) = \dfrac{1}{6} \sum_i{(t^{(i)}_0 - \dfrac{1}{6}\sum_i{t^{(i)}_0})^2}.
\end{equation}

Secondly, we used a double-phase exponential model~\cite{Bazin2009} with parameters: $f_0$, $t_0$, $f_m$, $\tau_{fall}$, and $\tau_{rise}$:
\begin{equation}
f^{(i)}(t)=f_m^{(i)} \cdot \dfrac{e^{-(t-t_0^{(i)})/\tau_{fall}}}{1 + e^{-(t-t_0^{(i)})/\tau_{rise}}} + f_0,
\end{equation}
where $i = 0..5$ is the passband number.

\begin{figure}[htb]
\includegraphics[width=1\textwidth]{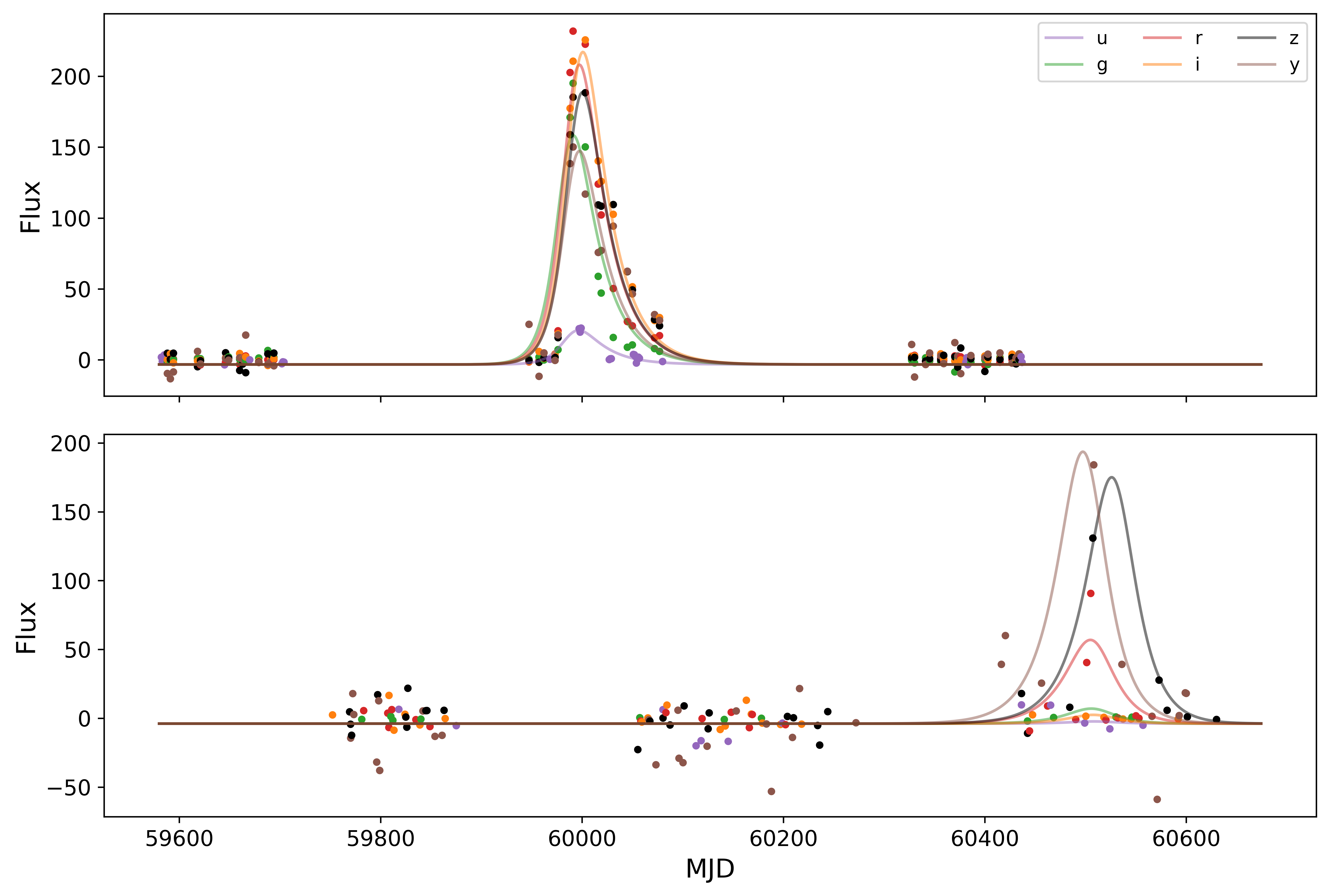}
\caption{The light curves data (dots) and fitted curves (Eqn.6, lines) for DDF (top) and WFD (bottom) samples.} \label{fig:fit}
\end{figure}

The examples of light curves fitted with Eqn.6 are shown in Fig.\ref{fig:fit} for DDF and WFD samples.
The fitted parameters, $f_m^{(i)}$, $\tau_{fall}$, and $\tau_{rise}$ were used as features for the classifier. The root-mean-square loss of the fits, $Bazin\_fitting\_loss$, was also added as a feature. From the Eqn.6 we derived the position of the peak maximum and found absolute magnitudes from the parametric curve fittings, $magnitude\_fitted\_i$. These fitted magnitude values are particularly useful for the incomplete light curves, where the flux maximum (the minimum of the absolute magnitude) was missed in the observations. We also calculated a range of features from the parametric fits for all 6 passbands: $log(f_m^{(i)})$, $m_{15}$ and $m_{-10}$ parameters~\cite{Taddia2015}, peak widths at three different levels of $0.2$, $0.5$ and $0.8$ of the flux peak maximum.

\subsubsection{Periodograms}
We used the Lomb-Scargle algorithm to extract periods from the non-periodic data \cite{Scargle1982}, which is a commonly-used statistical tool designed to detect periodic signals in unevenly-spaced observations. The periods have shown high permutation importance in the features selection process (Table 3). Adding $periods$ yield minor improvement of the model accuracy, which is probably not worth computational times (around two weeks for the test set on 24 CPUs). The dataset mostly consisted of the transients, and the other classes were well defined without Lomb-Scargle periods.

\subsubsection{Colours}
The astronomical colours were calculated from magnitudes as suggested in~\cite{Beare2014}, both from the measured detected magnitudes ($u\_r$, $u\_z$, etc.) and from the magnitude values found from the parametric curve fits (denoted as $fitted\_u\_z$).

\section{Machine learning method}
Boosted decision trees are, in general, excellent classifiers, commonly used for time series. Decision trees map input features to output classes using a series of decision rules~\cite{Quinlan1983}. The probabilities from each of the decision trees in the ensemble are averaged to produce a final probability. Averaging over a large number of classificators makes this method robust and increases the accuracy of predictions. Boosted decision trees have been successfully applied recently in photometric supernova classification~\cite{Lochner2016}.

In this paper, we use python boosted decision trees implementation, LightGBM~\cite{lightgbm}, with $5$ folds cross-validation, stratified by classes. We used different sets of features for the input of LightGBM classifier and selected the optimal features set based on the average 5-folds cross-validation scheme. The output was then tested on a large test dataset ($3479803$ light curves samples) provided in the challenge.

The hyper-parameters optimised for the LightGBM using cross-validation are listed in Table 2.
\begin{table}[htb]
\begin{center}
\begin{tabular}{|l|c|c|}
\hline
Parameter & Value & Description  \\
\hline
max\_depth& 3 & Maximum tree depth \\
min\_child\_samples & 50 & Minimum number of samples in a child (leaf).\\
num\_leaves& 5 & Maximum number of tree leaves\\
learning\_rate& 0.02 & Learning rate\\
max\_bin & 40 &  Max number of bins that feature values will be bucketed in\\
\hline
\end{tabular}
\end{center}
\label{table:params}
\caption{The hyper-parameters used for the LightGBM classifier.}
\end{table}

\section{Augmentations}
The training dataset is relatively small, and augmentations of the light curves were beneficial to reduce overfitting of the classifier.
The following augmentations were used:
\begin{itemize}
\item We first augmented train $10$x times using flux variation within a normal distribution of the corresponding \textit{flux-err} values provided. We also applied a similar approach to augment metadata for \textit{hostgal-photoz} values.

\item We introduced a small time shift to bring translation symmetry.

\item The train and test data had a different distribution with DDF data ratio of 30\% and 1\% in the train and test sets, respectively. To overcome over-fitting to the DDF samples, we down-sampled DDF in the current augmentation and re-selected features checking eli5 importance for WFD samples only.

\item Finally, we augmented train more, making 30x train from WFD samples and leaving DDF samples only in the original train fold. It gave further improvement to the model performance.
\end{itemize}

\section{Features selection}
The features selection is important to reduce overfitting, remove redundant features, and avoid confusing the classifier.
The following approaches were used to select the most relevant features for the task:
\begin{itemize}
\item We used an automatic tool, recursive feature elimination from sklearn library~\cite{rfe}, for the initial optimisation. This tool provides the first approximation of the good features set; however, it requires further manual fine-tuning.

\item We used permutation importance from \textit{eli5} package to rank features~\cite{eli5}, and then removed ones with negative or small importance. The permutation importance is computed as a decrease in the score when feature values are permuted (become noise). A list of the best-ranked features and their permutation importance gain coefficients is shown in Table 3.

\item We calculated the Pearson correlation between features and removed redundant one-by-one. Although decision tree models should be immune to correlated features, in practice removing highly correlated features improved the model performance.
    Such a manual approach with permutation importance worked out much better than the automatic feature selection.
\end{itemize}
\begin{table}
\begin{center}
\begin{tabular}{|l|c|c|c|c|c|}
\hline
Rank & Feature & Weight & Rank & Feature & Weight\\
\hline
1 & detected\_time\_diff & 0.2140$\pm$0.0187 & 51 & qso\_log\_chi2\_qsonu\_0 & 0.0031$\pm$0.0013\\
2 & hostgal\_photoz & 0.1119$\pm$0.0250 & 52 & m\_15\_1 & 0.0030$\pm$0.0015\\
3 & magnitude\_min & 0.0485$\pm$0.0147 & 53 & magnitude\_sum\_3 & 0.0030$\pm$0.0012\\
4 & Bazin\_fitting\_loss & 0.0280$\pm$0.0078 & 54 & median\_absolute\_deviation\_3 & 0.0030$\pm$0.0048\\
5 & magnitude\_mean & 0.0249$\pm$0.0102 & 55 & magnitude\_min\_2 & 0.0029$\pm$0.0017\\
6 & flux\_ratio\_square\_skew & 0.0219$\pm$0.0063 & 56 & supernova\_decay\_2 & 0.0028$\pm$0.0029\\
7 & fm\_0 & 0.0180$\pm$0.0022 & 57 & magnitude\_sum\_4 & 0.0028$\pm$0.0014\\
8 & median\_absolute\_deviation\_5 & 0.0167$\pm$0.0086 & 58 & r\_i & 0.0028$\pm$0.0007\\
9 & detected\_mean & 0.0161$\pm$0.0049 & 59 & qso\_log\_chi2\_qsonu\_2 & 0.0027$\pm$0.0010\\
10 & magnitude\_std & 0.0159$\pm$0.0036 & 60 & magnitude\_fitted\_5 & 0.0027$\pm$0.0018\\
11 & gauss\_s\_2 & 0.0138$\pm$0.0031 & 61 & flux\_skew\_1 & 0.0025$\pm$0.0024\\
12 & gauss\_error\_5 & 0.0133$\pm$0.0085 & 62 & autocorrelation & 0.0025$\pm$0.0030\\
13 & period & 0.0105$\pm$0.0039 & 63 & gauss\_s\_5 & 0.0024$\pm$0.0009\\
14 & norm\_flux\_max\_1 & 0.0102$\pm$0.0060 & 64 & magnitude\_mean\_0 & 0.0024$\pm$0.0012\\
15 & norm\_flux\_max\_5 & 0.0093$\pm$0.0043 & 65 & m\_15\_2 & 0.0023$\pm$0.0023\\
16 & median\_absolute\_deviation\_4 & 0.0089$\pm$0.0084 & 66 & magnitude\_mean\_4 & 0.0023$\pm$0.0033\\
17 & tau\_fall & 0.0088$\pm$0.0038 & 67 & qso\_log\_chi2\_qsonu\_5 & 0.0022$\pm$0.0013\\
18 & fitted\_g\_i & 0.0087$\pm$0.0016 & 68 & magnitude\_min\_fitted\_4 & 0.0021$\pm$0.0011\\
19 & g\_r & 0.0085$\pm$0.0016 & 69 & magnitude\_min\_5 & 0.0021$\pm$0.0013\\
20 & gauss\_s\_3 & 0.0074$\pm$0.0038 & 70 & log\_fm\_1 & 0.0020$\pm$0.0028\\
21 & peaks\_certain\_width & 0.0072$\pm$0.0045 & 71 & tau\_rise & 0.0020$\pm$0.0015\\
22 & CAR\_tau & 0.0067$\pm$0.0050 & 72 & fitted\_u\_r & 0.0020$\pm$0.0007\\
23 & log\_fm\_4 & 0.0064$\pm$0.0019 & 73 & m\_15\_4 & 0.0020$\pm$0.0017\\
24 & flux\_err\_min & 0.0064$\pm$0.0092 & 74 & magnitude\_sum\_0 & 0.0020$\pm$0.0010\\
25 & magnitude\_mean\_2 & 0.0063$\pm$0.0016 & 75 & fm\_1 & 0.0019$\pm$0.0023\\
26 & fitted\_g\_z & 0.0059$\pm$0.0009 & 76 & gauss\_s\_0 & 0.0019$\pm$0.0006\\
27 & flux\_diff & 0.0056$\pm$0.0014 & 77 & magnitude\_min\_fitted\_1 & 0.0019$\pm$0.0011\\
28 & gauss\_s\_4 & 0.0056$\pm$0.0025 & 78 & magnitude\_min\_4 & 0.0019$\pm$0.0012\\
29 & peaks\_estimated\_width & 0.0054$\pm$0.0023 & 79 & magnitude\_min\_fitted\_2 & 0.0018$\pm$0.0014\\
30 & magnitude\_sum\_2 & 0.0054$\pm$0.0019 & 80 & freq\_varrat\_4 & 0.0018$\pm$0.0026\\
31 & norm\_flux\_max\_0 & 0.0053$\pm$0.0018 & 81 & magnitude\_min\_0 & 0.0018$\pm$0.0008\\
32 & median\_absolute\_deviation\_2 & 0.0052$\pm$0.0025 & 82 & g\_i & 0.0018$\pm$0.0009\\
33 & gauss\_error\_2 & 0.0051$\pm$0.0026 & 83 & stetson\_k\_1 & 0.0016$\pm$0.0010\\
34 & m\_15\_5 & 0.0048$\pm$0.0026 & 84 & norm\_flux\_max\_4 & 0.0016$\pm$0.0005\\
35 & log\_fm\_5 & 0.0046$\pm$0.0027 & 85 & r\_z & 0.0015$\pm$0.0016\\
36 & norm\_flux\_max\_3 & 0.0045$\pm$0.0030 & 86 & fm\_3 & 0.0015$\pm$0.0009\\
37 & fitted\_i\_z & 0.0042$\pm$0.0018 & 87 & fitted\_r\_i & 0.0014$\pm$0.0009\\
38 & flux\_skew\_2 & 0.0042$\pm$0.0023 & 88 & magnitude\_mean\_3 & 0.0014$\pm$0.0008\\
39 & gauss\_error\_1 & 0.0041$\pm$0.0043 & 89 & g\_z & 0.0014$\pm$0.0011\\
40 & fitted\_r\_z & 0.0041$\pm$0.0018 & 90 & freq\_varrat\_1 & 0.0014$\pm$0.0018\\
41 & magnitude\_mean\_5 & 0.0039$\pm$0.0021 & 91 & magnitude\_min\_1 & 0.0013$\pm$0.0007\\
42 & fm\_5 & 0.0036$\pm$0.0018 & 92 & flux\_skew & 0.0013$\pm$0.0014\\
43 & gauss\_err\_4 & 0.0036$\pm$0.0047 & 93 & gauss\_error\_3 & 0.0013$\pm$0.0015\\
44 & fm\_2 & 0.0034$\pm$0.0016 & 94 & qso\_log\_chi2\_qsonu\_1 & 0.0012$\pm$0.0005\\
45 & i\_z & 0.0033$\pm$0.0014 & 95 & magnitude\_mean\_1 & 0.0012$\pm$0.0009\\
46 & norm\_flux\_max\_2 & 0.0032$\pm$0.0028 & 96 & CAR\_sigma & 0.0011$\pm$0.0017\\
47 & m\_15\_3 & 0.0032$\pm$0.0017 & 97 & log\_fm\_0 & 0.0010$\pm$0.0019\\
48 & gauss\_s\_1 & 0.0032$\pm$0.0019 & 98 & fitted\_g\_r & 0.0010$\pm$0.0007\\
49 & supernova\_decay\_3 & 0.0031$\pm$0.0013 & 99 & u\_r & 0.0009$\pm$0.0014\\
50 & magnitude\_sum\_5 & 0.0031$\pm$0.0019 & 100 & flux\_skew\_3 & 0.0009$\pm$0.0017\\
\hline
\end{tabular}
\end{center}
\label{table:eli}
\caption{A list of the best ranked features and their permutation importance coefficients.}
\end{table}

\section{Results}
The weighted logarithmic loss values for different sets of features used with LightGBM classifier are presented in Table 4 for 5-fold cross-validation and test data (the smaller the better).
\begin{table}[htb]
\begin{center}
\begin{tabular}{|l|c|c|c|}
\hline
Features set &  Augmentations & Log Loss, 5 folds & Log Loss, test \\
\hline
All features (167) & 10x augs & 0.52209 & 0.59910\\
Top 50 features & 10x augs&  0.52756 &  0.60863\\
Top 100 features & 10x augs& 0.51769 & 0.59770\\
Selected features (82)& 10x augs& 0.51588 & 0.59524 \\
Top 100 features & no augs& 0.44487 &  0.62834\\
Selected features (82)& no augs& 0.44036 & 0.63164\\
Selected features (82)& 30x augs& 0.51720 & 0.58788 \\
\hline
\end{tabular}
\end{center}
\label{table:results}
\caption{The model metric results for different sets of features.}
\end{table}

The evaluation metric, logarithmic loss, depended on the number of features included for the LightGBM classifier, with all (167) and too few features (top 50) giving the worst results. Including top $100$ features from the permutation importance ranking gave nearly optimal results. Removing the redundant features using Pearson correlation helped to improve the model accuracy further. The selected features set for the best model is listed in Table 5. Train set augmentation improved significantly model loss values between $0.63164$ and $0.58788$ for the test set. The cross-validation loss values were larger with augmentations, but it reduced overfitting giving better results on the test set.
\begin{table}[htb]
\begin{center}
\begin{tabular}{|c|}
\hline
Selected features   \\
\hline
$detected\_time\_diff$, $hostgal\_photoz$, $magnitude\_min$, $magnitude\_mean$,\\
$magniude\_std$, $magnitude\_min\_i, i= 0..5$, $magnitude\_mean\_i, i=2..5$, \\
$CAR\_tau$, $flux\_err\_min$, $norm\_flux\_max\_i, i = 0..5$,\\
$Bazin\_fitting\_loss$, $tau\_fall$, $tau\_rise$, $fm\_i, i=0..5$, $log\_fm\_i, i = 1, 4, 5$, \\
$flux\_ratio\_square\_skew$, $flux\_skew$, $flux\_skew\_1$, $flux\_skew\_2$, $flux\_diff$, \\
$median\_absolute\_deviation\_5$, $median\_absolute\_deviation\_2$, $detected\_mean$, \\
$peaks\_certain\_width$, $peaks\_estimated\_width$, $percent\_close\_to\_median\_2$, \\
$gauss\_s\_i, i = 0..5$, $gauss\_error\_i, i = 0..5$, $supernova\_decay\_3$, \\
$qso\_log\_chi2\_qsonu\_0$, $freq\_varrat\_1$, $freq\_varrat\_4$, $stetson\_k\_3$, \\
$magnitude\_fitted\_i, i = 1, 2, 4$, $autocorrelation$, $periods$\\
$g\_r$, $g\_i$, $r\_z$, $r\_i$, $g\_z$, $u\_r$, $i\_z$, \\
$fitted\_g\_r$, $fitted\_g\_i$, $fitted\_r\_z$, $fitted\_r\_i$, $fitted\_g\_z$, $fitted\_u\_r$, $fitted\_i\_z$ \\
\hline
\end{tabular}
\end{center}
\label{table:features}
\caption{The list of selected features set for the best model}
\end{table}

\begin{figure}[htb]
\includegraphics[width=1\textwidth]{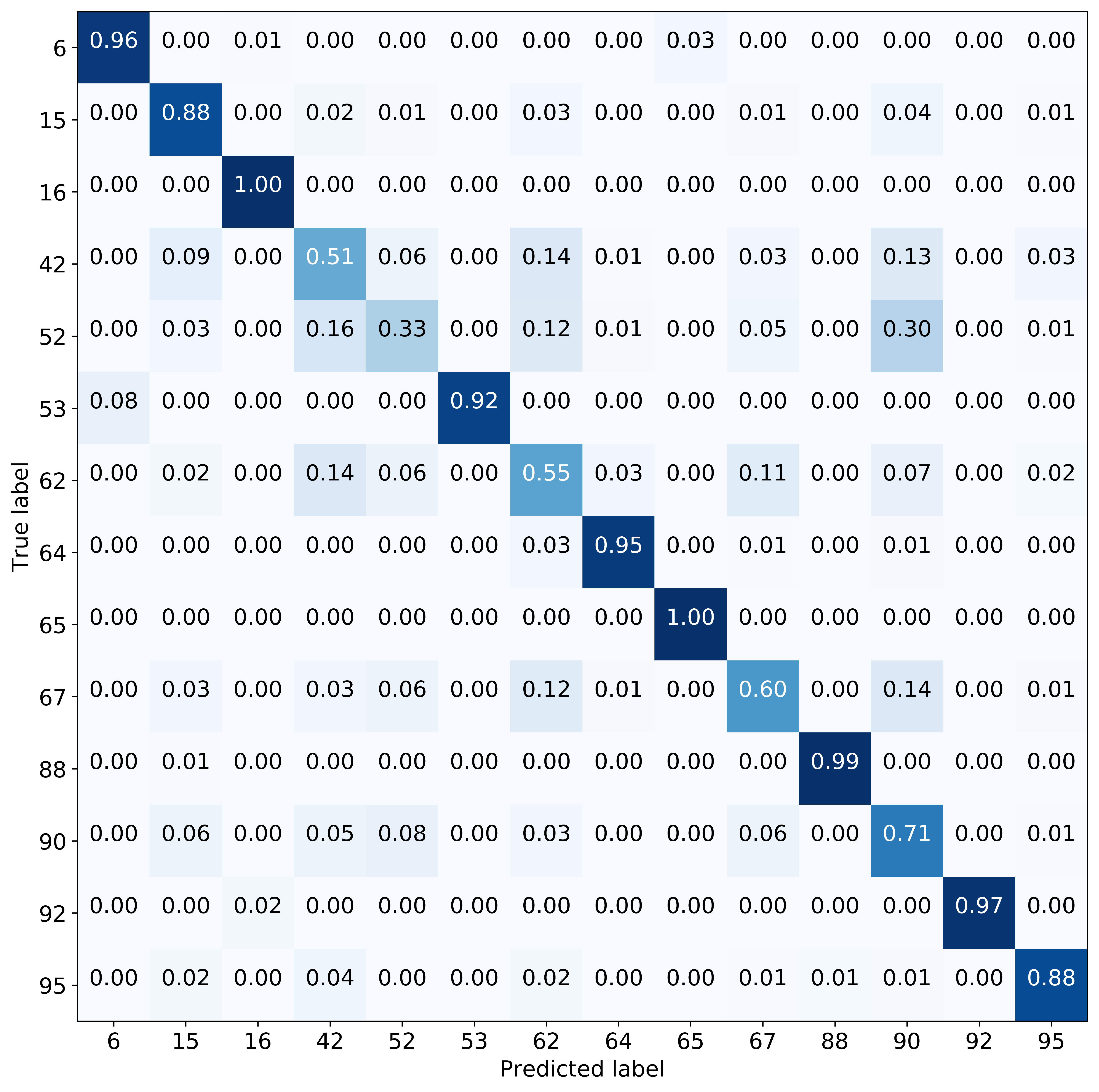}
\caption{The confusion matrix of the predictions for the best model with 5-folds cross-validation.} \label{fig:conf}
\end{figure}

The confusion matrix of the predictions for the best model with 5-folds cross-validation is shown in Fig.~\ref{fig:conf} The most problematic classes were supernova classes, showing the confusion of the classifier between different supernova types, especially for class 52 (SN Iax type). The other classes were well distinguished with the high accuracy of 88-100\% using the selected set of features. The model was tested on a large test dataset, giving one of the top results in the challenge.

\section{Discussion}
There was a range of other experiments that we tried, but it did not help to increase the accuracy of the model. Nevertheless, we list those unsuccessful attempts here, as we believe it might be useful information for others.

\subsubsection{Autoencoders}

We tried to use variational autoencoders to extract features from the light curves and use them for the classifier. We encoded the light curves in $43-50$ features; however, none of then gave improvement to the classifier when adding to the selected features set.

\subsubsection{Multiple parametric curve fittings}

We used several parametric fits from the papers of Bazin et.al.~\cite{Bazin2009}, Newling et.al.~\cite{Newling2011}, and Karpenka et.al.~\cite{Karpenka2012} to fit supernova rise and decay times. Using the aggregated rise and decay times gave the same accuracy than using individual values per passband. Adding parameters from several different fits together (i.e., form  Bazin et.al.~\cite{Bazin2009} and Karpenka et.al.~\cite{Karpenka2012}) did not bring improvement on classification, so that only single fit was sufficient. Based on the features selection process, we used Bazin et.al. fit~\cite{Bazin2009} for the final features set.

\section{Conclusion}
Here we present the model for the automatic photometric classification of the various astronomical objects. The model was developed in the context of the PLAsTiCC astrological challenge and achieved one of the top results in the competition.

Often, the solutions of machine learning competitions are based on extensive and diverse ensembles, test-time augmentation, and pseudo labeling, which is not always possible and feasible in real-life applications. We propose a solution based on the simple LightGBM model, feature extraction and augmentation.

We calculated and tested numerous features from the light curves and metadata. All features were ranked with permutation importance algorithm, and the most relevant were used for the classification (see Table 3). We used the boosted decision trees LightGBM model and 5-folds cross-validation. Too many features lead to overfitting of the classifier, while too few are insufficient to achieve a good model accuracy. Selecting around $100$ features seemed to be close to the golden middle for this task. Removing redundant features manually with Pearson correlation also improved the result. The features selection process is outlined in section 7.
Augmentations described in section 4 also helped to reduce overfitting and improve model accuracy. The final results and selected features are shown in Table 4 and Table 5, respectively.

We archived high accuracy of 88-100\% for classification for all astronomical object, except for the supernova classes: 42, 52, 62, 67, and 90. Supernova types classification accuracy varied between 33-71\%, with class 52 (SN Iax) being the most undefined. The confusion between different supernova types persisted, which indicates that other approaches would be desirable for these types. For instance, it could be beneficial to separate supernova types from the other objects and use an additional classifier for them and also introduce template fittings, similar to Type I~\cite{Guy2007}, for all supernova types.

\section{Acknowledgements}
The authors thank the challenge organisers, Dr Olivier Grellier for the introductory baseline~\cite{oliver-kernel}, Dr Grzegorz Sionkowski for suggesting one of the top features for the classifier, Kyle Boone, Dr Jean-François Puget (known as CPMP) and Open Data Science community (ods.ai) for useful discussions.

%
%

\bibliographystyle{unsrt}
\bibliography{plasticcbib}

\end{document}